\documentclass[a4paper, 11pt]{article}
\pdfoutput=1
\usepackage{jheppub}
\usepackage{afterpage}
\usepackage{multirow}
\usepackage{rotating}
\usepackage{epstopdf}

\title{Relevance of VHE Blazar Spectra Models with Axion-Like Particles}

\author{Hai-Jun Li}
\affiliation{Center for Advanced Quantum Studies, Department of Physics, Beijing Normal University, Beijing 100875, China}
\affiliation{Key Laboratory of Particle Astrophysics, Institute of High Energy Physics, Chinese Academy of Sciences, Beijing 100049, China}
\affiliation{School of Physics, University of Chinese Academy of Sciences, Beijing 100049, China}
\emailAdd{lihaijun@bnu.edu.cn}

\abstract{The oscillation of photons and axion-like particles (ALPs) in the astrophysical magnetic fields could modify the measured very high energy (VHE; $\mathcal{E}\gtrsim 100\, \rm GeV$) $\gamma$-ray spectra of the blazar sources. 
In this paper, we use the VHE $\gamma$-ray observations of the blazar Markarian 421 (Mrk~421) measured by MAGIC and \textit{Fermi}-LAT in 2017 with four phases to constrain the ALP.
We give the spectral energy distributions (SEDs) of these phases under the null and ALP hypotheses.
We also test the effects of the $\gamma$-ray blazar intrinsic spectra models on the ALP constraints. 
No significant relationship is confirmed between the ALP constraints and the model selections.
The 95\% $\rm C.L.$ combined constraints set by the single-model and multi-model scenarios on the ALP parameter space are roughly at the photon-ALP coupling $g_{a\gamma} \gtrsim 3\times 10^{-11} \rm \, GeV^{-1}$ for the ALP mass $1\times 10^{-8}\, {\rm eV} \lesssim m_a \lesssim 2\times 10^{-7}\, \rm eV$. 
}

\arxivnumber{2112.14145}

\begin{document}
\maketitle

\section{Introduction}

Axion-like particles (ALPs) are ultralight pseudo Nambu-Goldstone bosons with a two-photon vertex $g_{a\gamma}$ that predicted by several extensions of the Standard Model (SM), such as String Theory \cite{Svrcek:2006yi,Arvanitaki:2009fg}.
ALPs could account for the Dark Matter (DM) if they are produced non-thermally in the early Universe \cite{Preskill:1982cy, Sikivie:2009fv, Marsh:2015xka}.
They can also couple to the photons in the external magnetic field.
The effective Lagrangian for the photon-ALP interaction can be described by  \cite{Raffelt:1987im}
\begin{eqnarray}
\mathcal{L}_{\rm ALP}&=\frac{1}{2}\partial^\mu a\partial_\mu a - \frac{1}{2}m_a^2a^2 -\frac{1}{4}g_{a\gamma}aF_{\mu\nu}\tilde{F}^{\mu\nu},
\end{eqnarray}
where $a$ is the ALP field, $m_a$ is the ALP mass, $g_{a\gamma}$ is the coupling constant of the ALP and photons, and $F_{\mu\nu}$ and $\tilde{F}^{\mu\nu}$ are the electromagnetic field tensor and its dual tensor, respectively.
The coupling between the ALP and photons in the magnetic field environments would lead to the detectable photon-ALP oscillation effects, such as the excess of the soft X-rays from the center of galaxy clusters \cite{Conlon:2013txa, Angus:2013sua}, the anomalous stellar cooling \cite{Giannotti:2015kwo, Giannotti:2017hny}, the low energy electronic recoil event excess in XENON1T \cite{XENON:2020rca, DiLuzio:2020jjp}, and the TeV transparency of the Universe \cite{DeAngelis:2007dqd, Hooper:2007bq}.

Considering the very high energy (VHE) photons emitting from the astrophysical sources which are far from the Earth, the effect of photon-ALP oscillation could modify the measured photon spectra \cite{DeAngelis:2007dqd, Hooper:2007bq, Mirizzi:2007hr}.
In this sense, the photon-ALP oscillation could provide a natural way to drastically reduce the absorption of the VHE photons through the pair production processes that induced by the extragalactic background light (EBL) at energy above $100\, \rm GeV$ \cite{Simet:2007sa}.
The common mechanism is considering the photon-ALP conversion in the magnetic fields at the $\gamma$-ray source region and then further back-conversion in the magnetic field of the Milky Way \cite{Belikov:2010ma, Horns:2012kw}.
Thus, the $\gamma$-ray observations of the bright point sources, such as active galactic nuclei (AGNs), are the very suitable environments to constrain the ALP parameter space.
Many previous works have been performed to investigate the photon-ALP oscillation effect on the blazar spectra modifications \cite{Mirizzi:2009aj, DeAngelis:2011id, Dominguez:2011xy, Horns:2012kw, Mena:2013baa, Meyer:2013pny, Abramowski:2013oea,  Meyer:2014epa, Meyer:2014gta, Reesman:2014ova,TheFermi-LAT:2016zue, Galanti:2018upl, Galanti:2018myb,  Zhang:2018wpc, Liang:2018mqm,  Long:2019nrz, Libanov:2019fzq,   Pallathadka:2020vwu, Guo:2020kiq, Li:2020pcn, Li:2021gxs, Buehler:2020qsn, Cheng:2020bhr, Liang:2020roo, Long:2021udi, Davies:2020uxn, Batkovic:2021fzr}.

In Ref.~\cite{Guo:2020kiq}, we investigate the photon-ALP oscillation effect with the VHE $\gamma$-ray observations of the blazars PKS 2155$-$304 and PG 1553+113 measured by H.E.S.S. and \textit{Fermi}-LAT \cite{HESS:2016btr}.
In Ref.~\cite{Li:2020pcn}, we use the blazar Markarian 421 (Mrk~421) 4.5-year ($2008-2012$) $\gamma$-ray data measured by ARGO-YBJ and \textit{Fermi}-LAT \cite{Bartoli:2015cvo} to constrain the ALP.
The combined results of the ten phases of Mrk~421 give the ALP excluded region at $g_{a\gamma} \gtrsim 2\times 10^{-11} \rm \, GeV^{-1}$ for the ALP mass $5\times 10^{-10} \, {\rm eV} \lesssim m_a \lesssim 5\times 10^{-7}\, \rm eV$ at 95\% $\rm C.L.$ 
We find that the combined analysis can significantly improve constraint results.
In the above works, we use the single fixed $\gamma$-ray spectrum model to constrain the ALP, which is defined as ``single-model" scenario.
In Ref.~\cite{Li:2021gxs}, we use the VHE $\gamma$-ray data of Mrk~421 (fifteen phases; $2013-2014$) and PG 1553+113 (five phases) measured by MAGIC and \textit{Fermi}-LAT \cite{Acciari:2019zgl} to set constraints on the ALP parameter space. 
This give a slightly different 95\% $\rm C.L.$ excluded region at $8\times 10^{-9} \, {\rm eV} \lesssim m_a \lesssim 2\times 10^{-7}\, \rm eV$.
Compared with the VHE observations used in our previous studies, we choose the intrinsic spectrum with the minimum best-fit reduced $\chi^2$ under the null hypothesis in that work, which is defined as ``multi-model" scenario.

The main aim of this paper is to test the effects of $\gamma$-ray blazar intrinsic spectra models (single-model and multi-model) on the ALP constraints.
In this work, we use the VHE $\gamma$-ray spectra of Mrk~421 (four phases) measured by MAGIC and \textit{Fermi}-LAT in 2017 to constrain the ALP. 
The \textit{Fermi}-LAT data at lower energies ($0.1-100\, \rm GeV$) are collected during the common operation time.
We use these data to constrain the ALP and test the effects of intrinsic spectra models on the combined results.

The plan of this paper is as follows.
In Section~\ref{section_data}, we describe the $\gamma$-ray observations of the blazar Mrk~421 used in this work.
In Section~\ref{section_alp}, we briefly introduce the magnetic field environments of the photon-ALP beam propagating from the blazar region to the Milky Way.
The numerical results are presented and discussed in Section~\ref{section_result}.
The summary and conclusion are given in Section~\ref{section_sum}.

\section{Gamma-ray data of the blazar}
\label{section_data}

Blazars are recognized as a special class of radio-loud AGNs in the extragalactic VHE sky\footnote{\url{http://tevcat2.uchicago.edu}} \cite{Abdo:2009iq}. 
They host a super massive black hole ($10^6-10^9$ solar masses) at the center of the host galaxies with an elliptical morphology and eject a pair of relativistic plasma jets flowing in opposite directions  \cite{Kirk:1998kp}. 
The blazar jets axis is oriented along the line of sight of the observer at the Earth, which gives rise to strong relativistic beaming effects of the radiation. 
The origin and launching of the relativistic jets have not been fully understood, the common suggestion is that the jets are powered by the rotational energy of the black hole at the center.
Blazars can be divided into two categories, the BL Lacertae (BL Lac) objects and Flat Spectrum Radio Quasars (FSRQ) \cite{Urry:1995mg, Padovani:2017zpf}, depending on their observed characteristic optical spectral lines and their rest frame equivalent widths. 
BL Lac objects have no or weak spectral lines against the bright featureless optical band, while FSRQs exhibit strong broad emission lines.

\subsection{Mrk~421 observations by MAGIC and \textit{Fermi}-LAT}

Mrk~421 (RA=11$\rm ^h$4'27.31'', Dec=$38^\circ$12'31.8'', J2000), located at the redshift of $z_0$ = 0.031, is classified as the high-frequency peaked BL Lac object with the double-peak structure composed of many short flares in both the X-ray and $\gamma$-ray regions \cite{Abdo:2009iq}.
It is one of the most extensively studied and brightest extragalactic $\gamma$-ray sources in the VHE $\gamma$-ray sky.
Mrk~421 was first detected at VHE by the Whipple telescope in 1992 \cite{Punch:1992xw} and has been well regularly monitored by many Imaging Atmospheric Cherenkov Telescopes (IACTs) since then \cite{Aharonian:2002fk, Aharonian:2005ib, Acciari:2009sta, collaboration:2011esa,Abdo:2011zz}.

In Ref.~\cite{MAGIC:2021zhk}, they report the Mrk~421 observations from the multi-wavelength campaign measured from December 2016  to June 2017. 
The observations are provided by different instruments that covering the energy range from radio to VHE $\gamma$-ray. 
The X-ray observations are measured by \textit{Swift} and NuSTAR, while the $\gamma$-ray data are complemented by the observations from MAGIC, FACT, and \textit{Fermi}-LAT. 
In their observations, four phases MJD~57757 (January 4, 2017), MJD~57785 (February 1, 2017), MJD~57813 (March 1, 2017), and MJD~57840 (March 28, 2017) have a longer observing time compared with the majority of the measurements, which lasted about 50 minutes. 
In this paper, we select the  $\gamma$-ray observations of these four Mrk~421 phases measured by MAGIC and \textit{Fermi}-LAT for spectra analysis and set constraints on the ALP parameter space.
Furthermore, we study the effects of blazar intrinsic spectra models on the ALP constraints.

\subsection{Gamma-ray spectral energy distributions}
\label{section_data_2}

The broad-band spectral energy distribution (SED) of BL Lac objects can be characterized by the non-thermal radiation from the jet and shown as a distinct two-hump structure peaking components at low and high energies \cite{Muecke:2002bi}. 
The first one peak at infrared to X-ray frequencies is commonly explained as due to the synchrotron emissions from the ultrarelativistic electrons and/or positrons accelerated in the jet magnetic field \cite{Ghisellini:2017ico}.
The second hump component, peaking in the GeV to TeV energy regions, is most likely due to the electron inverse-Compton (IC) scattering off the synchrotron photons emitted by the same population of electrons.
This model is called as synchrotron self-Compton (SSC) model \cite{Chiang:1998wh, Petropoulou:2013lwa}. 
In this case, the presence of a sub-dominant hadronic component is also possible.
The peak frequency of the low energy component leads to a further subclassification of the BL Lac type objects.
This peak frequency ranges from IR–optical to UV–soft-X bands in low (LBL; $\nu_s<10^{14}\, \rm Hz$), intermediate (IBL; $10^{14}\, {\rm Hz}<\nu_s<10^{15}\, \rm Hz$), and high (HBL; $\nu_s>10^{15}\, \rm Hz$) synchrotron peaked BL Lacs \cite{Abdo:2009iq}.
The most extreme high-frequency peaked BL Lacs (EHBL; $\nu_s>10^{17}\, \rm Hz$) showing IC peaks in the energy range above $100\, \rm GeV$ \cite{Bonnoli:2015yia}.
However, there is currently a lack of detailed investigation of such an extreme peak exceeding TeV energies and the hard intrinsic spectrum at sub-TeV energies.
This feature shows that there is a hard accelerated particle spectrum, most of which is carried by the particles with the highest energy.

In this work, we mainly focus on the blazar spectra of the VHE component.
The $\gamma$-ray blazar intrinsic spectra $\Phi_{\rm int} (\mathcal{E})$ adopted here can be described by the simple and smooth concave functions with three to five free parameters, defining as power law with exponential cut-off (EPWL; three parameters), power law with superexponential cut-off (SEPWL; four parameters), log parabola (LP; four parameters), and log parabola with exponential cut-off (ELP; five parameters).
The functional expressions can be described by \cite{Li:2021gxs}
\begin{eqnarray}
{\rm EPWL}: \Phi_{\rm int} ( \mathcal{E} ) &=& \mathcal{A}_0\left(\frac{\mathcal{E}}{\mathcal{E}_0}\right)^{-\Gamma}\exp\left(-\frac{\mathcal{E}}{\mathcal{E}_c}\right),\\
{\rm SEPWL}: \Phi_{\rm int} ( \mathcal{E} ) &=& \mathcal{A}_0\left(\frac{\mathcal{E}}{\mathcal{E}_0}\right)^{-\Gamma}\exp\left(-\left(\frac{\mathcal{E}}{\mathcal{E}_c}\right)^d\right),\\
{\rm LP}: \Phi_{\rm int} ( \mathcal{E} ) &=& \mathcal{A}_0\left(\frac{\mathcal{E}}{\mathcal{E}_0}\right)^{-\Gamma-b \log\left(\frac{\mathcal{E}}{\mathcal{E}_0}\right)},\\
{\rm ELP}: \Phi_{\rm int} ( \mathcal{E} ) &=& \mathcal{A}_0\left(\frac{\mathcal{E}}{\mathcal{E}_0}\right)^{-\Gamma-b \log\left(\frac{\mathcal{E}}{\mathcal{E}_0}\right)}\exp\left(-\frac{\mathcal{E}}{\mathcal{E}_c}\right),
\end{eqnarray}
where $\mathcal{A}_0$ is the normalization constant, $\mathcal{E}_0$ is the normalization energy, $\Gamma$ is the spectral index, $\mathcal{E}_c$,  $b$, and $d$ are free parameters. 
For EPWL and SEPWL, we adopt $\mathcal{E}_0 = 1\, \rm GeV$, while for LP and ELP, the parameter $\mathcal{E}_0$ is also taken as a free parameter.

\begin{figure}[!htbp]
\centering
   \includegraphics[width=1\textwidth]{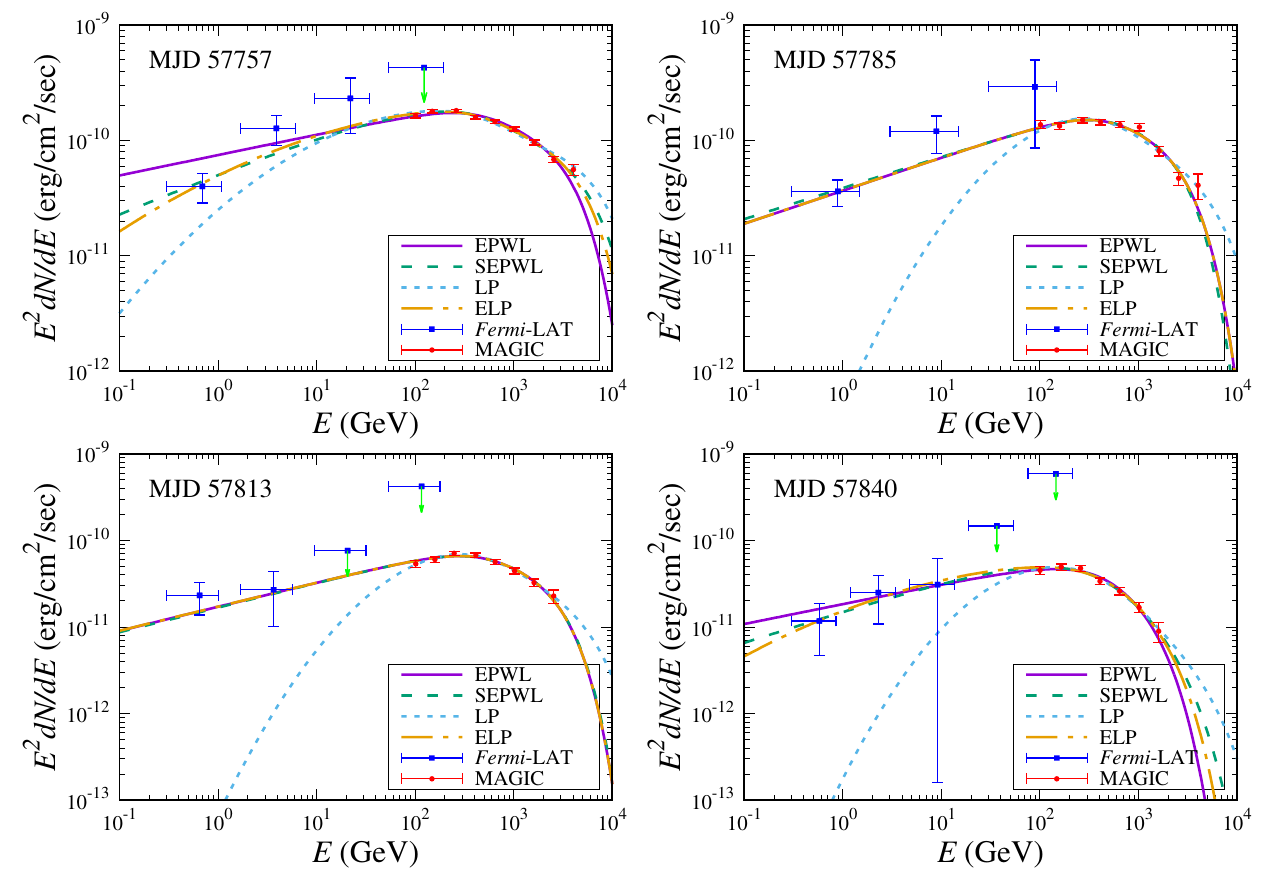}
  \caption{The best-fit $\gamma$-ray SEDs of Mrk~421 under the null hypothesis in the phases MJD~57757 (top left), MJD~57785 (top right), MJD~57813 (bottom left), and MJD~57840 (bottom right). The different lines represent the spectra with the four intrinsic spectra models. The values of best-fit $\chi^2$ are listed in Table~\ref{tab_1}. The experimental data of MAGIC and \textit{Fermi}-LAT are taken from Ref.~\cite{MAGIC:2021zhk}.
}
  \label{fig_dnde_1}
\end{figure}

An important effect for the VHE photons crossing in the extragalactic space is the EBL absorption effect through the pair production process $\gamma_{\rm VHE} + \gamma_{\rm BG} \to e^+ + e^-$. 
This effect can be characterized by the optical depth $\tau$ that depending on the redshift of the source and the distribution of the EBL.
In our analysis, we take the EBL model from Ref.~\cite{Franceschini:2008tp}.
Then we can derive the expected $\gamma$-ray spectrum under the null hypothesis using the EBL absorption factor $e^{-\tau}$
\begin{eqnarray}
\Phi_{\rm w/o \; ALP} ( \mathcal{E} ) = e^{-\tau} \Phi_{\rm int} ( \mathcal{E} ),
\end{eqnarray}
with the intrinsic spectrum $\Phi_{\rm int}(\mathcal{E})$.

In the fitting, the $\chi^2$ value can be defined as \cite{Acciari:2019zgl}
\begin{eqnarray}
\chi^2 = \sum_{i=1}^{N} \left(\frac{\Phi(\mathcal{E}_i) - \tilde{\phi}_i}{\delta_i}\right)^2,
\label{chi2}
\end{eqnarray}
where $N$ is the point number of the spectrum, $\Phi(\mathcal{E}_i)$ is the expected photon flux, $\tilde{\phi}_i$ is the detected photon flux, and $\delta_i$ is the corresponding uncertainty of the measurement.

Then we can derive the best-fit spectra for the four phases of Mrk~421 under the null hypothesis with four intrinsic spectra models.
The best-fit SEDs for these phases are shown in Figure~\ref{fig_dnde_1}.
The values of best-fit $\chi^2$ and $\chi^2/\rm d.o.f.$ are also listed in Table~\ref{tab_1}.
Since the small uncertainties of the phase and the parameter number of EPWL is less than other models, the phase MJD~57757 under EPWL model shows a large value of ${\chi}_{\rm w/oALP}^2/\rm d.o.f.$
We also notice that the ${\chi}_{\rm w/oALP}^2/\rm d.o.f.$ values  of the model LP are large for the phases MJD~57757 and MJD~57785.
However, this spectrum model is suitable for the situation with high EBL absorption, such as the blazar PG 1553+113 considered in Ref.~\cite{Guo:2020kiq}.

\begin{table}
\centering
\begin{tabular}{lcccccr}
\hline\hline
Phase   &        Model  &  ${\chi}_{\rm w/oALP}^2$ &  ${\chi}_{\rm w/oALP}^2/\rm d.o.f.$  &  $\chi^2_{\rm min}$    &   eff. $\rm d.o.f.$    &   $\Delta\chi^2$    \\
\hline
MJD~57757    &  EPWL    &  28.759   &3.195&   14.253  &   4.65   &     10.547\\
($4^{\rm th}$ January 2017)             &  SEPWL    &  13.247     &1.656&   5.669    &   4.48   &     10.279\\ 
             &  LP     &  23.917   &2.990&   11.196  &   4.76   &     10.720\\
             &  ELP     &  13.738     &1.963&   7.528    &   4.64   &     10.532\\ 
\hline
MJD~57785     &  EPWL    &  13.326   &1.481 &   5.286    &   4.44   &     10.216\\
($1^{\rm st}$ February 2017)              &  SEPWL    &  13.326   &1.666 &   5.097    &   4.52   &     10.343\\ 
              &  LP     &  36.544   &4.568 &   15.489  &   4.58   &     10.437\\
              &  ELP     &  13.325   &1.904 &   4.705    &   4.74   &     10.688\\ 
\hline
MJD~57813    &  EPWL    &   4.286    &0.612 &   1.538    &   4.27   &     9.946\\
($1^{\rm st}$ March 2017)             &  SEPWL    &   4.275    &0.713 &   1.195    &   3.99   &     9.495\\ 
             &  LP     &   9.719    &1.620 &   8.930    &   3.96   &     9.447\\
             &  ELP     &   4.286    & 0.857&   1.538    &   3.98   &     9.479\\ 
\hline
MJD~57840    &  EPWL    &   5.961    &0.852 &   0.743    &   4.00   &     9.511\\
($28^{\rm th}$ March 2017)            &  SEPWL    &   4.489    &0.748 &   0.643    &   4.02   &     9.544\\ 
             &  LP     &   9.441    &1.574 &   2.291    &   3.90   &     9.349\\
             &  ELP     &   4.838    &0.968 &   0.756    &   3.95   &     9.430\\ 
 \hline\hline
\end{tabular}
\caption{List of the values of best-fit ${\chi}_{\rm w/oALP}^2$ under the null hypothesis for the four phases of Mrk~421 with the different intrinsic spectra models.  The values of the minimum best-fit $\chi^2_{\rm min}$ in the $m_a-g_{a\gamma}$ plane under the ALP hypothesis are shown. The effective $\rm d.o.f.$ of the TS distributions and the values of $\Delta{\chi}^2$ corresponding to 95\% $\rm C.L.$ are also listed.}
\label{tab_1}
\end{table}

\begin{table}\centering
\begin{tabular}{lccccr}
\hline\hline
Phase    &     $B_0^{\rm jet}$($10^{-2}\, \rm G$)    &   $r_{\rm VHE}$($10^{17}\, \rm cm$)   &    $n_0^{\rm jet}$($10^3\, \rm cm^{-3}$)  & $\delta_{\rm D}$  \\
\hline
MJD~57757  &   6.1     &  2.5   &   7.534  &  25\\
MJD~57785  &   7.0     &  2.5   &   6.036  &  25\\
MJD~57813  &   6.1     &  4.125  &   1.665  &  25\\
MJD~57840  &   10.0   &  3.325  &   1.519  &  25\\
 \hline\hline
\end{tabular}
\caption{The BJMF model parameters in the four phases of Mrk~421. These values are derived from Ref.~\cite{MAGIC:2021zhk}.}
\label{tab_2}
\end{table}

\section{Axion-like particles effect}
\label{section_alp}

In this section, we briefly introduce the magnetic field environments of the photon-ALP oscillation.
For the VHE photons propagating from the source to our Earth, we consider the photon-ALP oscillation effect in three regions, including the source region, the extragalactic space, and the Milky Way \cite{DeAngelis:2007dqd,Hooper:2007bq}.
The final photon survival probability can be described by \cite{DeAngelis:2011id,Meyer:2014epa}
\begin{eqnarray}
\mathcal{P}_{\gamma\gamma}={\rm Tr}\left(\left(\rho_{11}+\rho_{22}\right)\mathcal{T}(s)\rho(0)\mathcal{T}^\dagger(s)\right),
\label{pgg}
\end{eqnarray}
with $\rho_{ii}={\rm diag}(\delta_{i1},\delta_{i2},0)$, where $\mathcal{T}(s)$ is the whole transfer matrix for the propagation distance $s$, $\rho(0)$ and $\rho(s)=\mathcal{T}(s)\rho(0)\mathcal{T}^\dagger(s)$ are the initial and final photon-ALP beam density matrices, respectively.

\begin{figure}[!htbp]
\centering
  \includegraphics[width=1\textwidth]{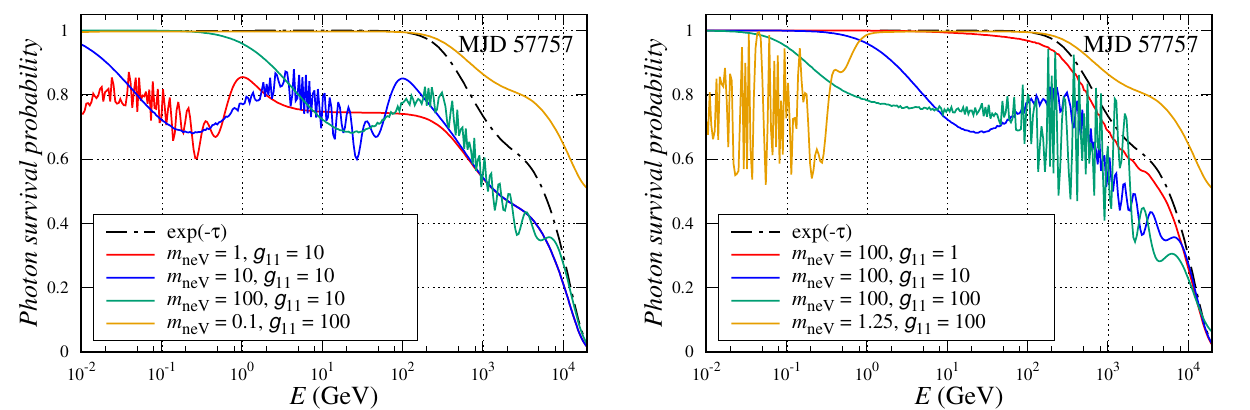}
  \caption{The photon survival probability for the phase Mrk~421 MJD~57757. The black dotted dashed line represents the photon survival probability without the photon-ALP oscillation. The solid lines represent the photon survival probability induced by the effects of the EBL absorption and the photon-ALP oscillation for several typical points on the ALP parameter space. 
The EBL model used here is taken from Ref.~\cite{Franceschini:2008tp}. 
The BJMF model parameters are given by Table \ref{tab_2}. Here $m_{\rm neV} \equiv m_a/1 \, \rm {neV} $ and $g_{11}\equiv g_{a\gamma}/10^{-11} \, \rm{GeV}^{-1}$.}
  \label{fig_pa}
\end{figure}

For the source region of Mrk 421, we mainly consider the photon-ALP oscillation in the blazar jet magnetic field (BJMF).
Following Refs.~\cite{Galanti:2018upl, Li:2020pcn}, this jet transverse magnetic field $B^{\rm jet}(r)$ can be described by \cite{Begelman:1984mw, Ghisellini:2009wa}
\begin{eqnarray}
B^{\rm jet}(r)=B_0^{\rm jet}\left(\frac{r}{r_{\rm VHE}}\right)^{\eta_{\rm jet}},
\label{bjet}
\end{eqnarray}
where $B_0^{\rm jet}$ is the core magnetic field strength at $r_{\rm VHE}$, $r_{\rm VHE}$ is the distance between the central black hole and the emission region of the VHE photons, and $\eta_{\rm jet}=-1$.
The density distribution of the electron $n_{\rm el}^{\rm jet}(r)$ is \cite{OSullivan:2009dsx}
\begin{eqnarray}
n_{\rm el}^{\rm jet}(r)=n_0^{\rm jet}\left(\frac{r}{r_{\rm VHE}}\right)^{\beta_{\rm jet}},
\end{eqnarray}
where $n_0^{\rm jet}$ is the electron density at $r_{\rm VHE}$ and $\beta_{\rm jet}=-2$.
The value of $r_{\rm VHE}\sim R_{\rm VHE}/\theta_{\rm jet}$ depends on the radius of the VHE emitting plasma blob $R_{\rm VHE}$ and $\theta_{\rm jet}$, which is the angle between the line of sight and jets axis. 
The photon energy in the laboratory frame $\mathcal{E}_L$ is determined by the energy in the co-moving frame as $\mathcal{E}_L = \mathcal{E}_j\cdot\delta_{\rm D}$ with the Doppler factor $\delta_{\rm D}$.
Above BJMF parameters listed in Table~\ref{tab_2} can be obtained from the fit with the SSC model in Ref.~\cite{MAGIC:2021zhk}.

When the photon-ALP beam propagates in the extragalactic space, the photon-ALP oscillation may occur in the extragalactic magnetic field.
In Ref.~\cite{Galanti:2018upl}, this field is modeled as a domain-like network with the common benchmark value $1 \, \rm nG$.
The current limit of the extragalactic magnetic field on the largest cosmological scale of ${\cal O} (1)$ Mpc is $10^{- 7} \, {\rm nG} \lesssim {B}_{\rm ext} \lesssim 1.7 \,  {\rm nG}$ \cite{Durrer:2013pga, Pshirkov:2015tua, Ade:2015cva}.
The value of this magnetic field is very small and can not be clearly determined, so the photon-ALP oscillation in the extragalactic magnetic field is neglected in our analysis.
The main effect on the VHE photons is still the EBL absorption through the pair production process (see Section~\ref{section_data_2}).

We also take into account the photon-ALP oscillation in the Galactic magnetic field of the Milky Way. 
This Galactic magnetic field model can be found in Refs.~\cite{Jansson:2012pc, Jansson:2012rt}, which is composed of the disk and the halo components, both parallel to the Galactic plane, and the out-of-plane component as ``X-field" at the Galactic center.
The latest version about this model can be found in Refs.~\cite{Unger:2017kfh, Planck:2016gdp}.

\section{Numerical results and discussions}
\label{section_result}
 
\subsection{Photon survival probability}
 
Using the magnetic field environments considered in Section~\ref{section_alp}, we can derive the whole transfer matrix as
\begin{eqnarray}
\mathcal{T}(s)=\mathcal{T}(s_3)_{\rm MW}\times\mathcal{T}(s_2)_{\rm EXT}\times\mathcal{T}(s_1)_{\rm BJ},
\end{eqnarray}
where $\mathcal{T}(s_1)_{\rm BJ}$, $\mathcal{T}(s_2)_{\rm EXT}$, and $\mathcal{T}(s_3)_{\rm MW}$ are the transfer matrices of the blazar jet, the extragalactic space, and the Milky Way, respectively.
Putting this into Eq.~(\ref{pgg}), we can derive the final survival probability of the VHE photons at the Earth.

An example for the photon survival probability of the phase Mrk~421 MJD~57757 is plotted in Figure~\ref{fig_pa}.
The black dotted dashed line represents the photon survival probability without the photon-ALP oscillation. 
The solid lines represent the photon survival probability induced by the effects of the EBL absorption and the photon-ALP oscillation for several typical points (for $m_{\rm neV}$=1, 10, 100 and $g_{11}$ = 1, 10, 100; with red, blue, and green lines, respectively) on the ALP parameter space. 
We can find that the EBL absorption effect would significantly suppress the SED of Mrk~421 in the form of $e^{-\tau}$ at energy above $100 \, \rm GeV$ . 
Compared with this effect, the photon-ALP oscillation may affect the SED start from the low energy regions.
For the same ALP mass $m_a$, the intensity of the photon-ALP oscillation changes dramatically with the coupling constant $g_{a\gamma}$.
Additionally, we also show the case that for some ALP parameters the survival probability in presence of ALP is consistently larger than the standard case with only the EBL absorption effect in VHE regions (solid yellow lines).
 
\subsection{Chi-square statistic}

Considering the photon-ALP oscillation effect, we can derive the expected $\gamma$-ray spectrum under the ALP hypothesis using the final survival probability of the photons $P_{\gamma\gamma}$ in Eq.~(\ref{pgg}), which reads
\begin{eqnarray}
\Phi_{\rm w \; ALP} ( \mathcal{E} ) = \mathcal{P}_{\gamma\gamma} \Phi_{\rm int} ( \mathcal{E} ),
\end{eqnarray}
with the intrinsic spectrum $\Phi_{\rm int}(\mathcal{E})$. 
Putting this into Eq.~(\ref{chi2}), we can derive the value of $\chi_{\rm w \; ALP}^2$ under the ALP hypothesis.

\begin{figure}[!htbp]
\centering
  \includegraphics[width=0.95\textwidth]{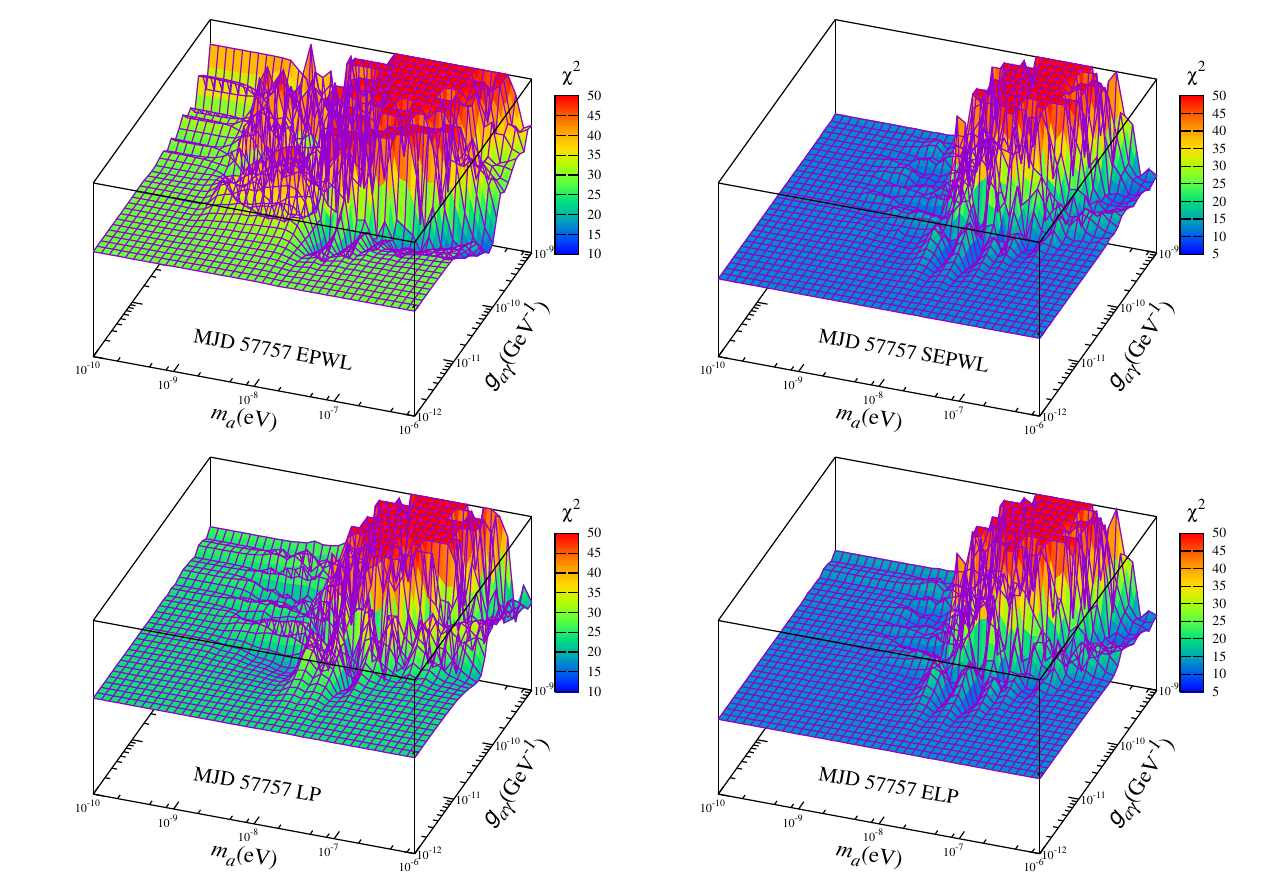}
   \includegraphics[width=0.95\textwidth]{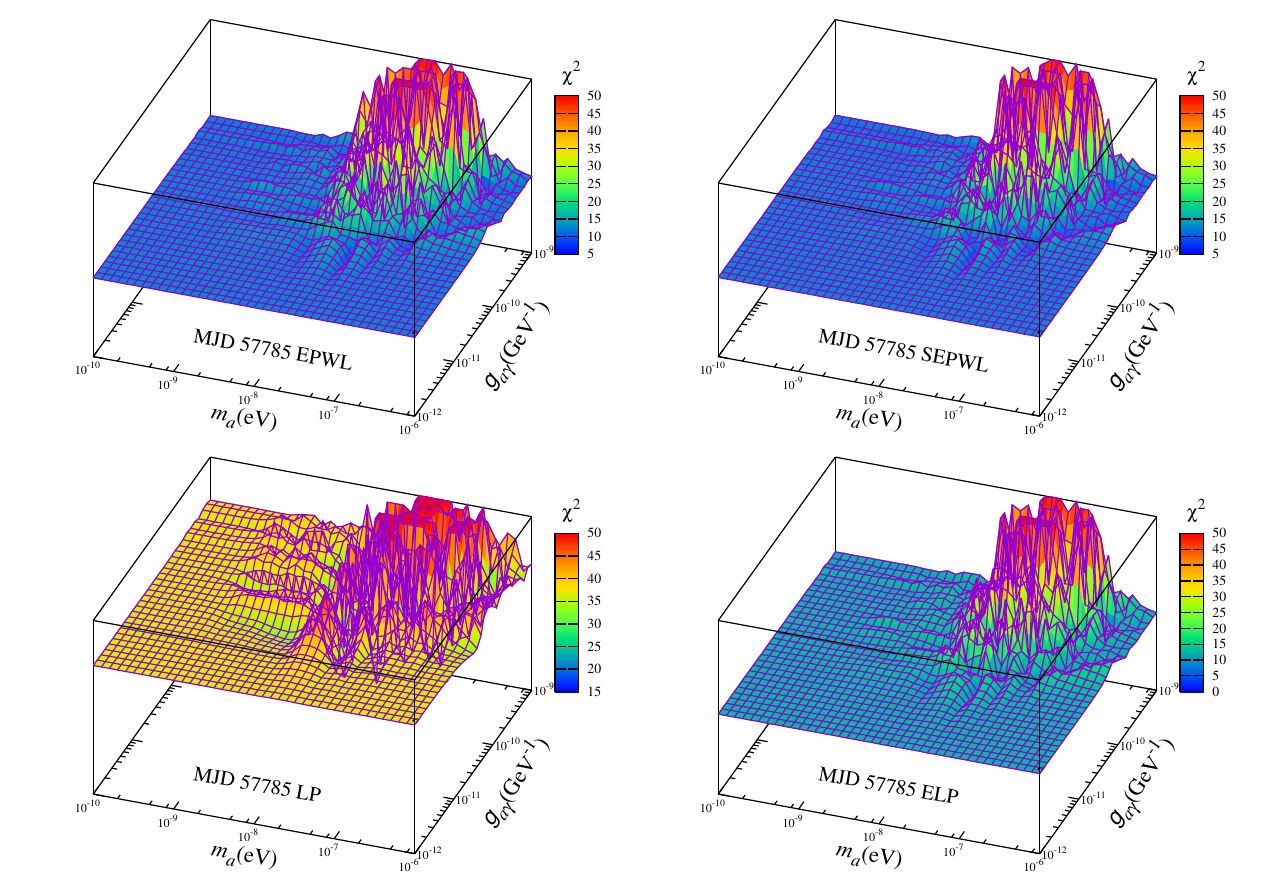}
  \caption{The distributions of $\chi_{\rm w \; ALP}^2$ on the $m_a-g_{a\gamma}$ parameter space for the two phases (MJD~57757 and MJD~57785) of Mrk~421 under the four intrinsic spectra models (EPWL, SEPWL, LP, and ELP).}
  \label{fig_chi2_1}
\end{figure}

\begin{figure}[!htbp]
\centering
  \includegraphics[width=0.95\textwidth]{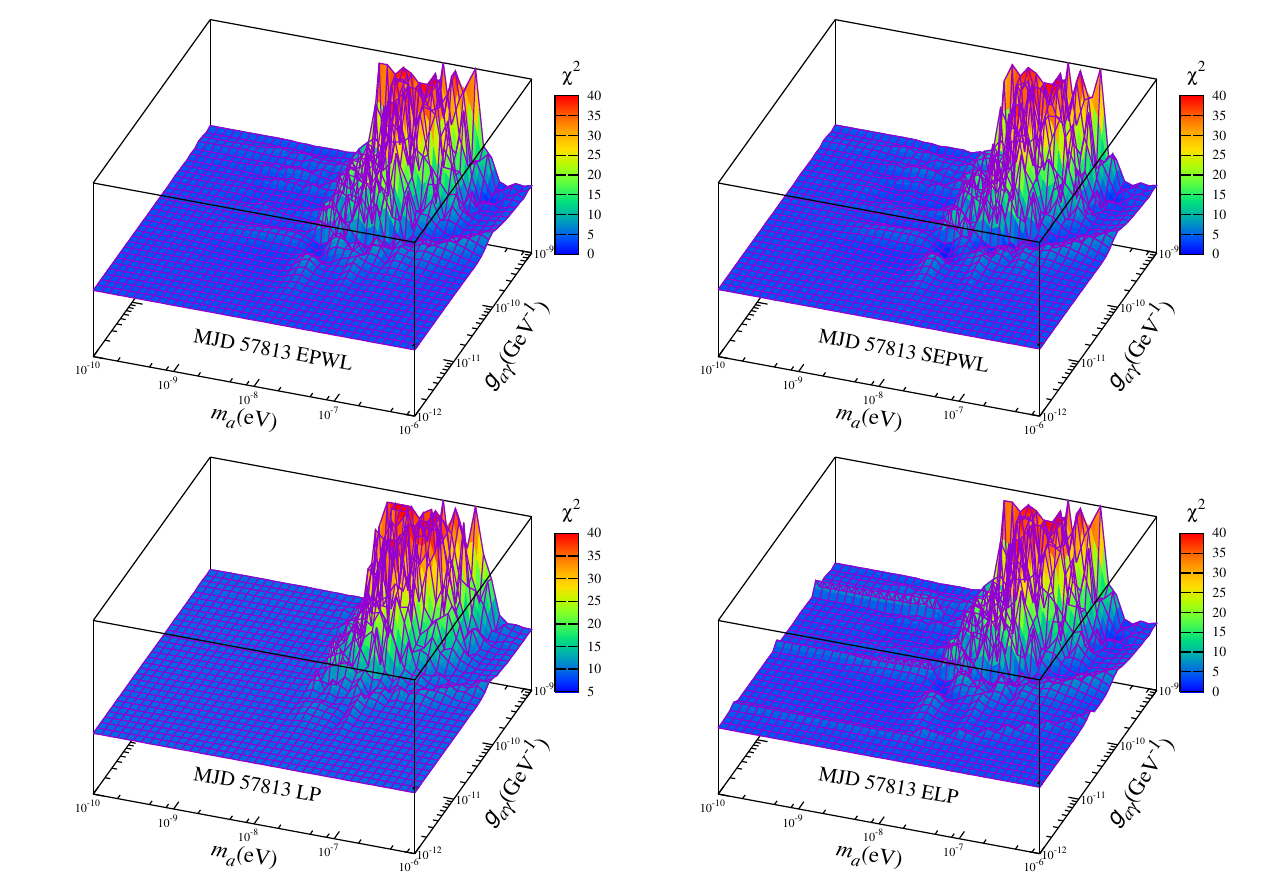}
   \includegraphics[width=0.95\textwidth]{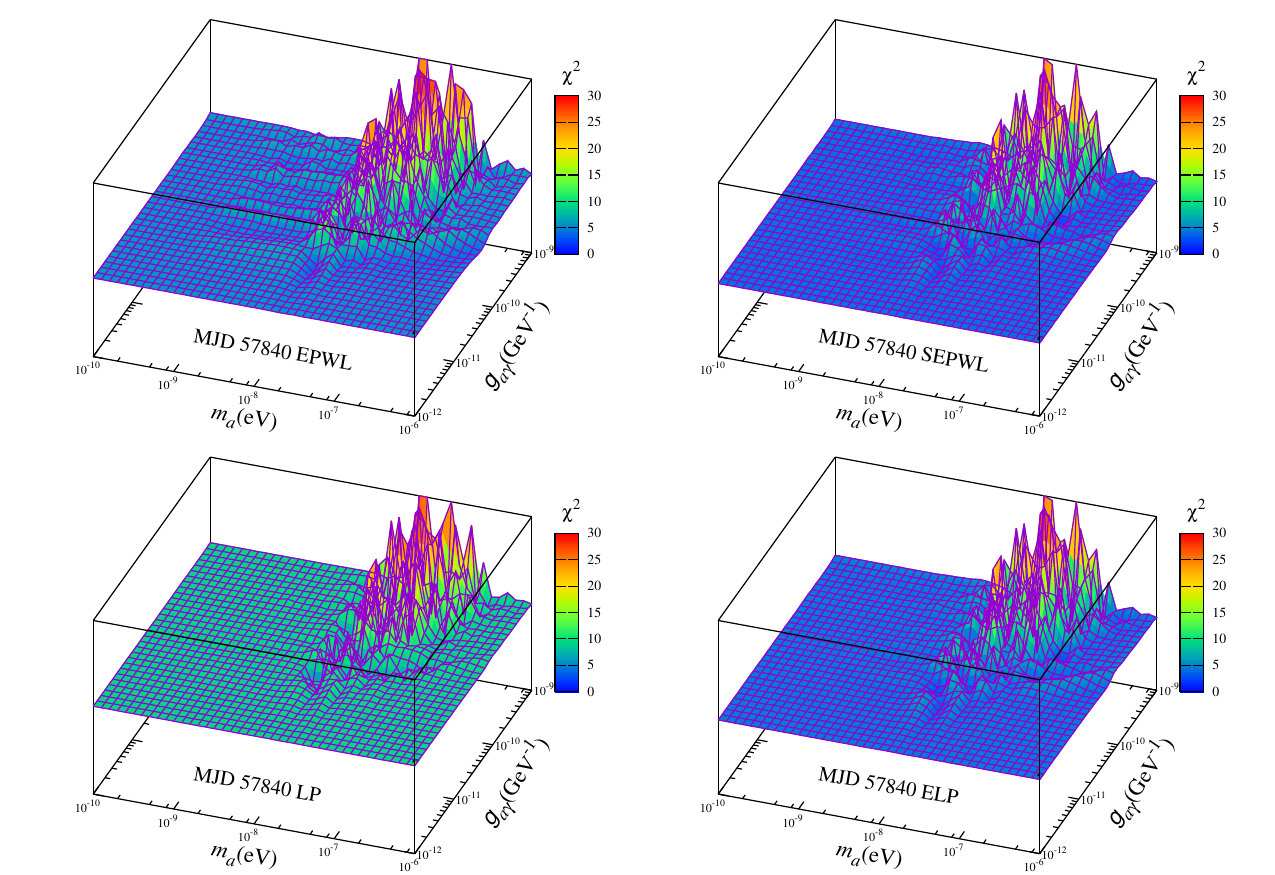}
  \caption{Same as Figure~\ref{fig_chi2_1} but for the two phases MJD~57813 and MJD~57840 of Mrk~421 under the four intrinsic spectra models.}
  \label{fig_chi2_2}
\end{figure}

\begin{figure}[!htbp]
\centering
  \includegraphics[width=1\textwidth]{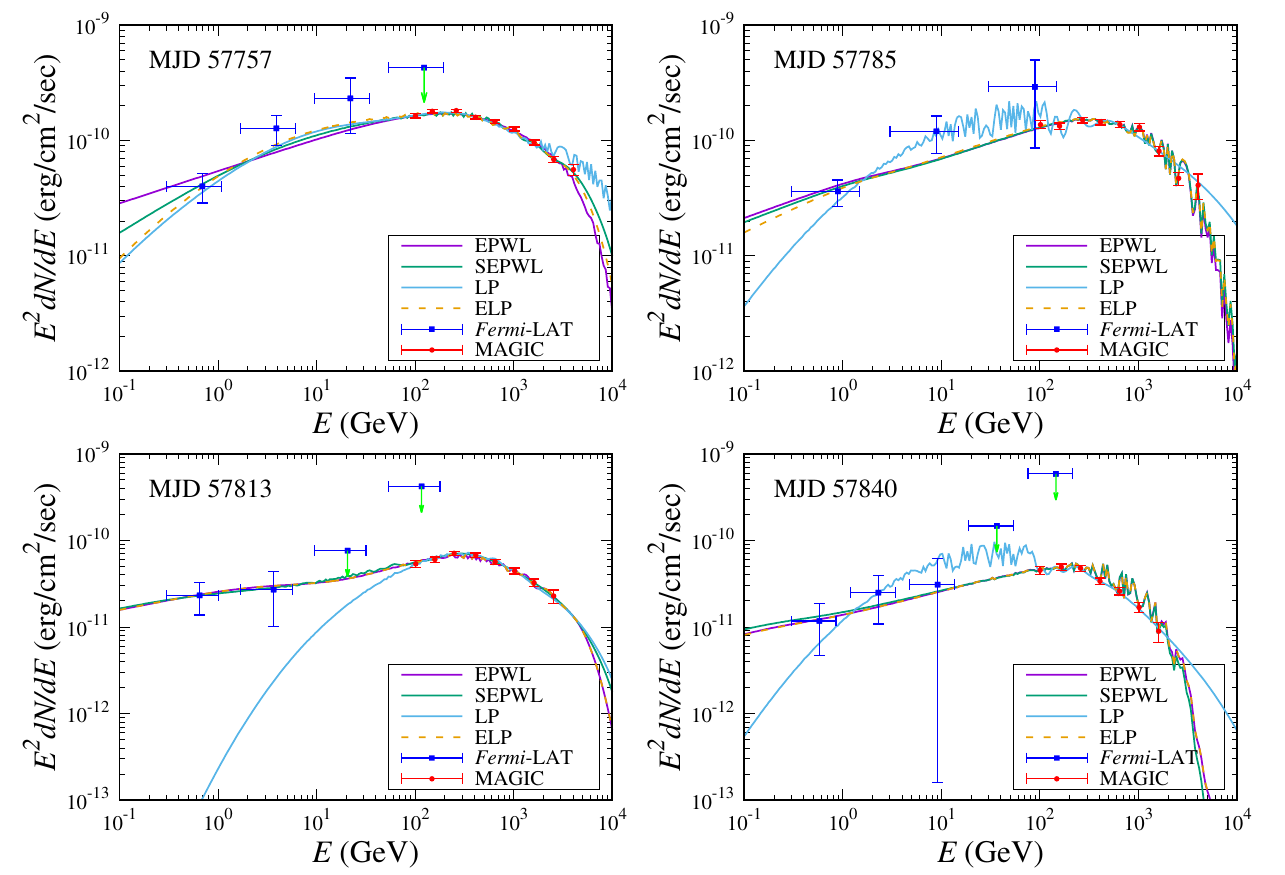}
  \caption{The best-fit $\gamma$-ray SEDs of Mrk~421 under the ALP hypothesis in the phases MJD~57757 (top left), MJD~57785 (top right), MJD~57813 (bottom left), and MJD~57840 (bottom right). The different lines represent the spectra with the four intrinsic spectra models. The values of best-fit $\chi^2$ are listed in Table~\ref{tab_1}. The BJMF model parameters are given by Table \ref{tab_2}. The experimental data of MAGIC and \textit{Fermi}-LAT are taken from Ref.~\cite{MAGIC:2021zhk}.
}
  \label{fig_dnde_2}
\end{figure}

Then we can derive the best-fit $\chi_{\rm w \; ALP}^2$ for each ALP parameter set in the $m_a-g_{a\gamma}$ plane.
The distributions of $\chi_{\rm w \; ALP}^2$ under the ALP hypothesis in the four phases are shown in Figure~\ref{fig_chi2_1} and \ref{fig_chi2_2} with the BJMF parameters listed by Table~\ref{tab_2}.
The minimum best-fit values of ${\chi}_{\rm min}^2$ in the whole parameter space for these phases are given in Table~\ref{tab_1}.
We also give the minimum best-fit $\gamma$-ray SEDs for the phases of Mrk~421 under the ALP hypothesis in Figure~\ref{fig_dnde_2}.

\subsection{Bounds on the ALP parameter space}

Following Refs.~\cite{TheFermi-LAT:2016zue, Li:2020pcn}, we simulate 400 sets of the VHE photon observations generated by Gaussian samplings in the pseudo-experiments to derive the test statistic (TS) distribution.
It can be defined by the best-fit ${\widehat{\chi}}^2$ under both the null and ALP hypotheses for each observation set, ${\rm TS} ={\widehat{\chi}_{\rm null}}^2 - {\widehat{\chi}_{\rm w \; ALP}}^2$, which obeys the non-central $\chi^2$ distribution with the non-centrality $\lambda$ and the effective degree of freedom ($\rm d.o.f.$). 
We take the TS distribution from the pseudo-experiments as the approximation of the ALP hypothesis and use it to derive the value of $\Delta{\chi}^2$ at the particular confidence level.
Then we can derive the threshold value $\chi_{\rm th}^2$ to set the ALP constraint in the $m_a-g_{a\gamma}$ plane, which is defined as $\chi_{\rm th}^2 = \chi_{\rm min}^2 + \Delta{\chi}^2$ with the minimum best-fit ${\chi}_{\rm min}^2$ under the ALP hypothesis.

\begin{figure}[!htbp]
\centering
  \includegraphics[width=1\textwidth]{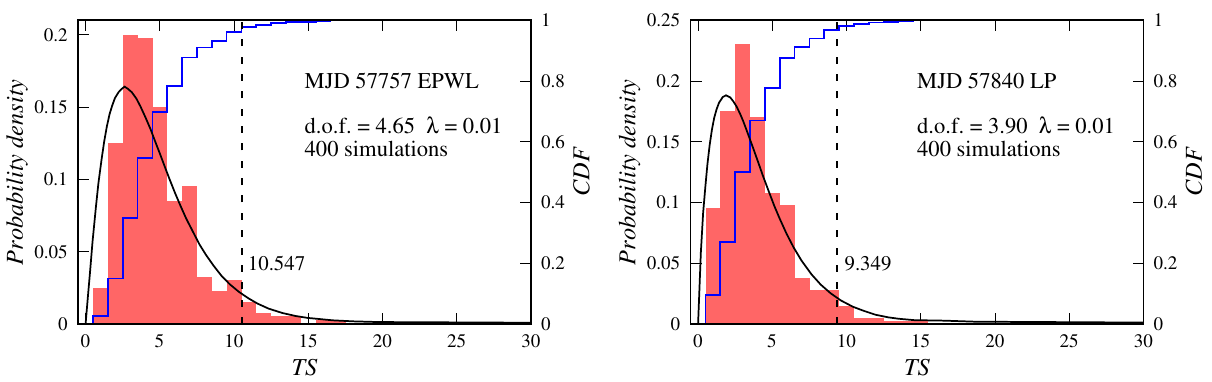}
  \caption{The TS distributions of the phases MJD~57757 under EPWL model (left) and MJD~57840 under LP model (right). The black lines represent the fitted non-central $\chi^2$ distributions.  The blue lines represent the CDF of the TS distributions.
}
  \label{fig_ts}
\end{figure}

We find that the non-centralities of all the TS distributions for the four phases under the different intrinsic spectra models are about 0.01, which indicates that the standard $\chi^2$ function can fit the distribution well.
We show the effective $\rm d.o.f.$ of the distributions in Table~\ref{tab_1}. 
In Figure~\ref{fig_ts}, we plot the TS distributions for the phases MJD~57757 under EPWL model and MJD~57840 under LP model. 
The black and blue lines represent the fitted non-central $\chi^2$ distributions and the cumulative distribution functions (CDF), respectively.
With these distributions, we can derive the values of $\Delta{\chi}^2$ corresponding to the 95\% $\rm C.L.$, which are also shown in Table~\ref{tab_1}.

\begin{figure}[!htbp]
\centering
  \includegraphics[width=1\textwidth]{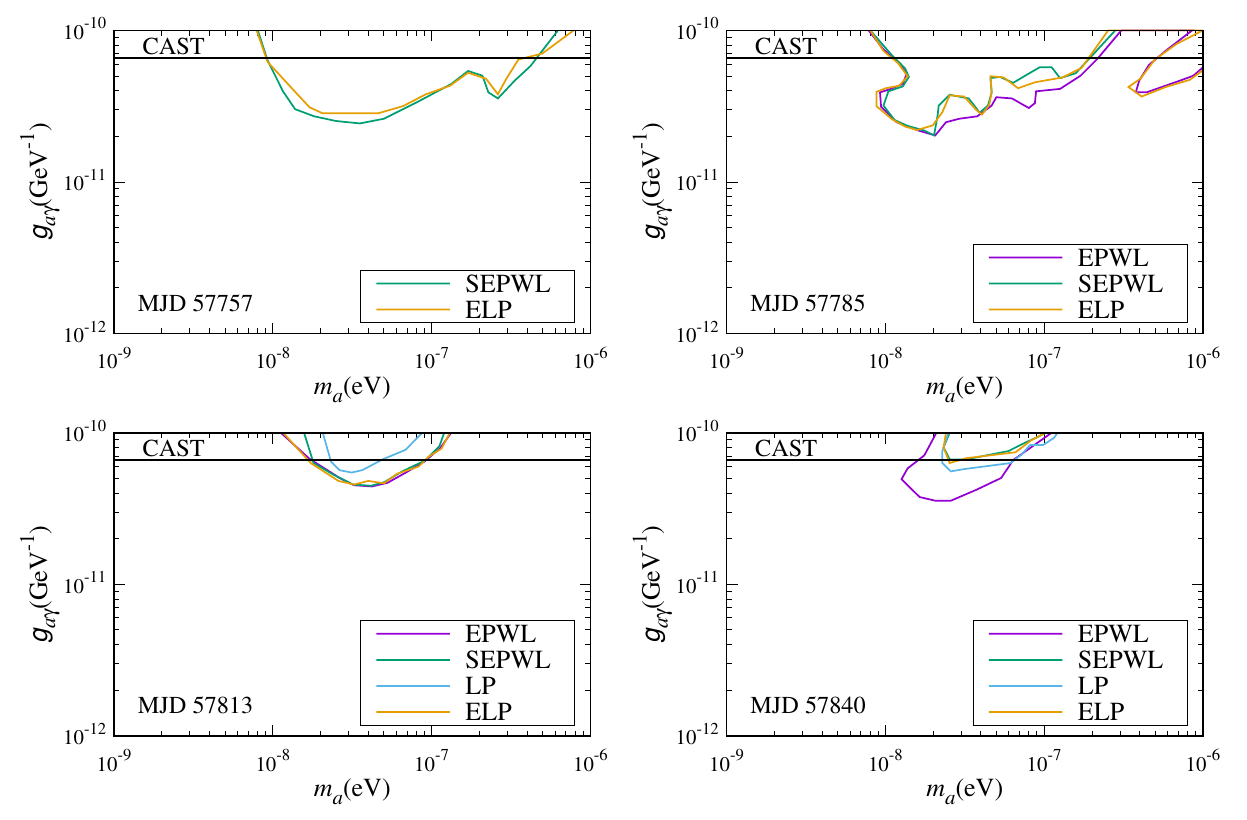}
  \caption{The excluded regions at 95\% $\rm C.L.$ set by the Mrk~421 observations of MAGIC and \textit{Fermi}-LAT with the four phases MJD~57757 (top left), MJD~57785 (top right), MJD~57813 (bottom left), and MJD~57840 (bottom right) under the different intrinsic spectra models. The contours represent the 95\% $\rm C.L.$ excluded regions. The black horizontal line represents the upper limit set by CAST \cite{Anastassopoulos:2017ftl} with $g_{a\gamma} < 6.6 \times 10^{-11}\, \rm GeV^{-1}$. The $\chi_{\rm w \; ALP}^2$ distributions on the $m_a-g_{a\gamma}$ parameter space for these phases are shown in Figure~\ref{fig_chi2_1} and \ref{fig_chi2_2}. 
}
  \label{fig_compare}
\end{figure}

Using the values of $\Delta{\chi}^2$ for these Mrk~421 phases under the different intrinsic spectra models, we expect to give the 95\% $\rm C.L.$ excluded regions on the ALP parameter space.
In Figure~\ref{fig_compare}, we show the ALP constraints with the Mrk~421 observations by MAGIC and \textit{Fermi}-LAT.
However, we find that not all the situations of these phases can be individually used to set the 95\% $\rm C.L.$ constraint, such as the phase MJD~57757 under EPWL and LP models, and the phase MJD~57785 under LP model. 
In these situations, the values of the best-fit ${\chi}_{\rm w/oALP}^2$ under the null hypothesis are less than the corresponding threshold values at 95\% $\rm C.L.$
For the excluded regions with the same phase shown in Figure~\ref{fig_compare}, we can not find a significant relationship between the ALP constraints and the model selections.
In the phase MJD~57840, the EPWL model would give a more stringent excluded region on the ALP parameter space.

\begin{figure}[!htbp]
\centering
  \includegraphics[width=0.95\textwidth]{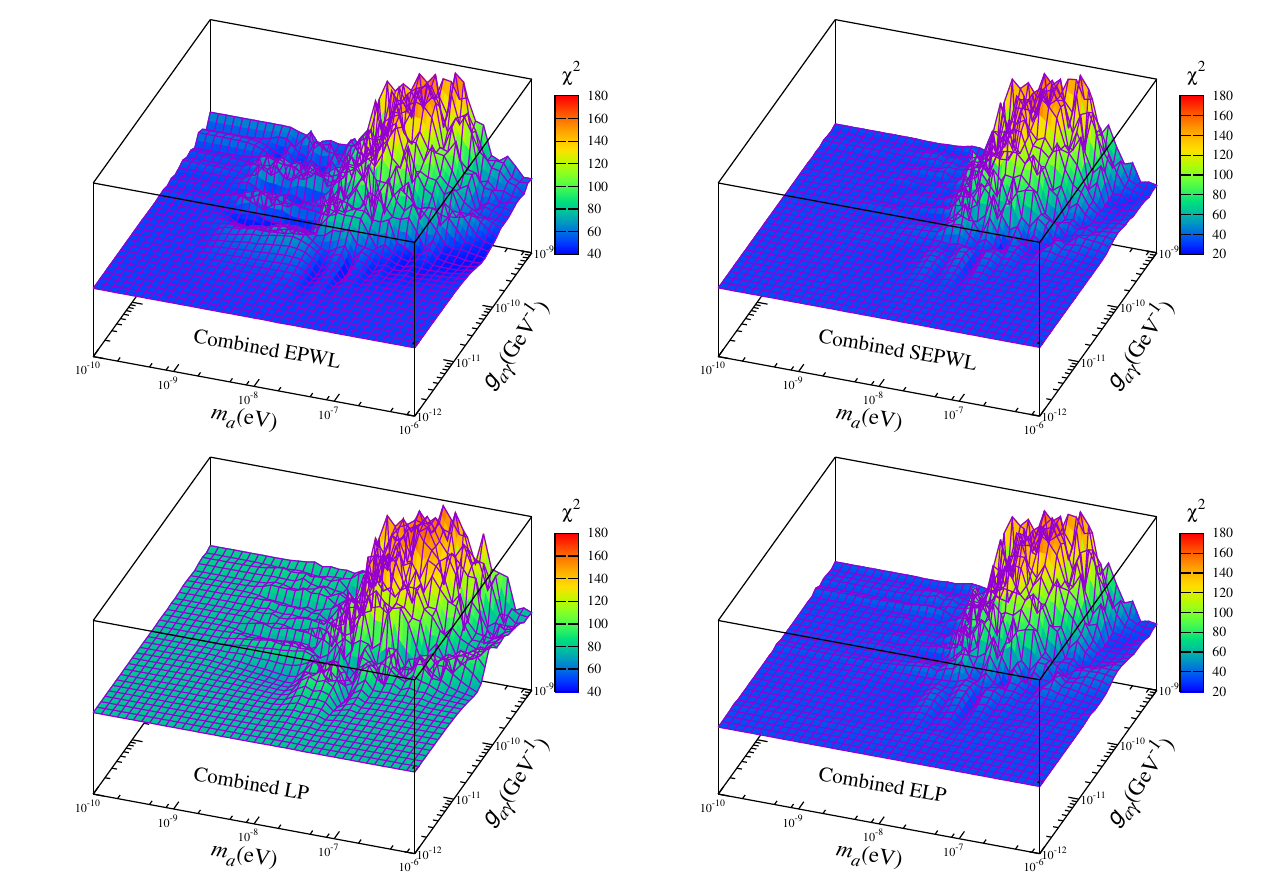}
  \caption{The distributions of $\chi_{\rm w \; ALP}^2$ on the $m_a-g_{a\gamma}$ parameter space for the single-model EPWL (top left), SEPWL (top right), LP (bottom left), and ELP (bottom right) combined results.}
  \label{fig_chi2_combined}
\end{figure}

\subsubsection{Combined constraints}

\begin{figure}[!htbp]
\centering
  \includegraphics[width=0.75\textwidth]{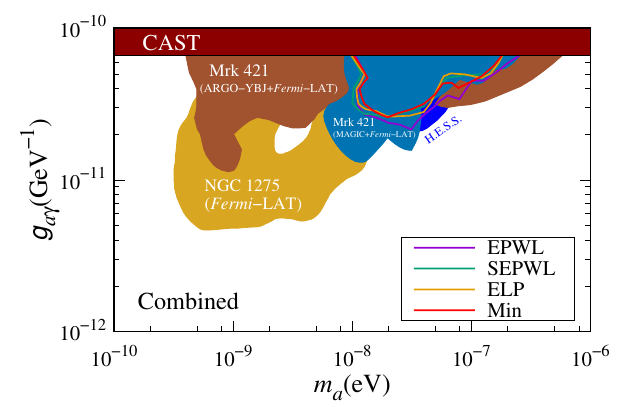}
  \caption{The excluded regions at 95\% set by the Mrk~421 observations of MAGIC and \textit{Fermi}-LAT with the four phases combined. The spectra models (``EPWL", ``SEPWL", and ``ELP") represent the combined results set by the single-model scenarios. The ``Min" represents the combined result set by the multi-model scenario. The black horizontal line represents the upper limit set by CAST \cite{Anastassopoulos:2017ftl} with $g_{a\gamma} < 6.6 \times 10^{-11}\, \rm GeV^{-1}$. The $\chi_{\rm w \; ALP}^2$ distributions on the $m_a-g_{a\gamma}$ parameter space for the single-model and multi-model combined results are shown in Figure~\ref{fig_chi2_combined} and \ref{fig_chi2_combined_min}, respectively. We also show the results of the PKS 2155$-$304 observation by H.E.S.S. \cite{Abramowski:2013oea}, the NGC 1275 observation by \textit{Fermi}-LAT \cite{TheFermi-LAT:2016zue}, and the Mrk~421 observations by ARGO-YBJ+MAGIC+\textit{Fermi}-LAT \cite{Li:2020pcn, Li:2021gxs} for comparisons.
}
  \label{fig_compare_2}
\end{figure}

Following Ref.~\cite{Li:2020pcn}, we also show the combined constraints at 95\% $\rm C.L.$ for these phases together.
This combined approach is necessary for the multi-phase analysis and could give a more reliable implication.
In Figure~\ref{fig_chi2_combined}, we give the $\chi_{\rm w \; ALP}^2$ distributions of the combined phases under the single intrinsic spectrum model (single-model).
We also give the excluded regions at 95\% $\rm C.L.$ set by the single-model scenario in Figure~\ref{fig_compare_2}.
For the spectra models except LP, we could set the 95\% $\rm C.L.$ combined constraint on the ALP parameter space.

As considered in Ref.~\cite{Li:2021gxs}, we also give the combined result under the intrinsic spectrum that chosen with the minimum best-fit reduced $\chi^2$ under the null hypothesis (multi-model).
From Table~\ref{tab_1}, this multi-model scenario is that SEPWL for phase MJD~57757, EPWL for phase MJD~57785, EPWL for phase MJD~57813, and SEPWL for phase MJD~57840.
In Figure~\ref{fig_compare_2}, the ``Min" represents the excluded regions at 95\% $\rm C.L.$ set by the multi-model scenario.
Compared with the single-model results, the excluded regions are roughly coincide with that derived from the multi-model scenario.
We have the conclusion that no significant relationship is confirmed between the ALP constraints and the model selections.
The 95\% $\rm C.L.$ combined constraints set by the single-model and multi-model scenarios on the ALP parameter space are roughly at the photon-ALP coupling $g_{a\gamma} \gtrsim 3\times 10^{-11} \rm \, GeV^{-1}$ for the ALP mass $1\times 10^{-8}\, {\rm eV} \lesssim m_a \lesssim 2\times 10^{-7}\, \rm eV$, which is approximately coincident with that derived from the $2013-2014$ observations of Mrk~421 in Ref.~\cite{Li:2021gxs}.

\begin{figure}[!htbp]
\centering
  \includegraphics[width=0.7\textwidth]{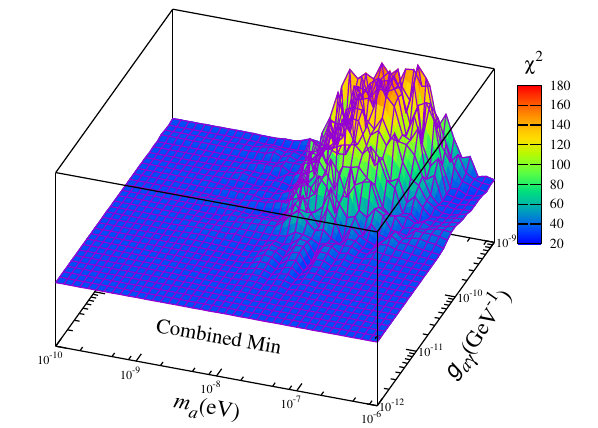}
  \caption{The distribution of $\chi_{\rm w \; ALP}^2$ on the $m_a-g_{a\gamma}$ parameter space for the multi-model (SEPWL for phase MJD~57757, EPWL for phase MJD~57785, EPWL for phase MJD~57813, and SEPWL for phase MJD~57840) combined result.}
  \label{fig_chi2_combined_min}
\end{figure}

\section{Summary and conclusion}
\label{section_sum}

In this paper, we use the VHE $\gamma$-ray observations of the blazar Mrk~421 measured by MAGIC and \textit{Fermi}-LAT during the common operation time to constrain the ALP, which covers the four activity phases in 2017.
In the analysis, we adopt the four $\gamma$-ray blazar intrinsic spectra models EPWL, SEPWL, LP, and ELP. 
We find that not all the situations of these phases can be individually used to set the 95\% $\rm C.L.$ constraint in the $m_a-g_{a\gamma}$ plane.
We also test the effects of the intrinsic spectra models on the ALP constraints. 
Compared with the single-model and multi-model scenarios, very similar results are obtained for the 95\% $\rm C.L.$ excluded regions on the ALP parameter space with the photon-ALP coupling $g_{a\gamma} \gtrsim 3\times 10^{-11} \rm \, GeV^{-1}$ for the ALP mass $1\times 10^{-8}\, {\rm eV} \lesssim m_a \lesssim 2\times 10^{-7}\, \rm eV$. 
We find that no significant relationship is confirmed between the ALP constraints and the model selections.
However, the multi-model scenario is essential for the multi-phase analysis to reduce the uncertainties from the $\gamma$-ray observations and the BJMF model.

The future astrophysical observations of the VHE $\gamma$-rays, such as LHAASO \cite{Cao:2010zz}, HERD \cite{Huang:2015fca}, CTA \cite{Acharya:2013sxa}, TAIGA-HiSCORE \cite{Kuzmichev:2018mjq}, and GAMMA~400 \cite{Egorov:2020cmx}, will provide more opportunities to constrain the ALP in this astrophysical approach.
Additionally, we expect that ALP could be directly detected in the future experiments, such as IAXO \cite{IAXO:2019mpb}, ALPS~II \cite{Spector:2019ooq}, STAX \cite{Capparelli:2015mxa}, and ABRACADABRA \cite{Kahn:2016aff}.

\section*{Acknowledgments}
The author would like to thank Axel Arbet-Engels and David Paneque for providing the $\gamma$-ray observations of Mrk~421 by MAGIC and \textit{Fermi}-LAT in the common operation time.
This work is partly supported by the National Key R\&D Program of China (Grant No.~2016YFA0400200) and the National Natural Science Foundation of China (Grants No.~U1738209 and No.~11851303) in IHEP, and partly supported by the National Natural Science Foundation of China (Grants No.~11775025 and No.~12175027) in BNU.


\bibliographystyle{JHEP}
\bibliography{references}

\end{document}